\documentclass[fleqn,twoside]{article}
\usepackage{amsmath}
\usepackage{espcrc2,psfig,epsfig}
\usepackage{graphicx}
\usepackage[figuresright]{rotating}

\title{Improved low $Q^2$ model for neutrino and electron nucleon
 cross sections in few GeV region}

\author{Arie Bodek\address[roc]{Department of Physics and Astronomy,
             University of Rochester,
             Rochester, New York 14618,  USA},
        Inkyu Park\addressmark[roc],
	and Un ki Yang
	\address[Chi]{Enrico Fermi Institute, University of Chicago,
	Chicago,USA}}
    
\begin{document}

\begin{abstract}
We present an improved model on neutrino- and electron-nucleon 
scattering cross sections using effective leading order PDFs
with a new scaling variable $\xi_w$. Non-perturbative QCD effects 
at low $Q^2$ are separately treated for $u$ and $d$ valence quarks,
and sea quarks. The improved model uses all inelastic charged lepton
$F_2$ data (SLAC/BCDMS/NMC/HERA), and photoproduction data
on proton and deuterium. In this way, we obtain an improved model
which describes all inelastic scattering charged lepton data,
JLAB resonance data, and neutrino data. This improved model
is expected to be better for neutrino oscillation experiments
at few GeV region.
(Presented by Un Ki Yang at NuInt04,   
Mar. 2004, Laboratori Nazionali del Gran Sasso -INFN - Assergi, 
Italy)
\vspace{1pc}
\end{abstract}

\maketitle

Recent discovery of neutrino oscillation from atmospheric neutrino
experiment~\cite{ATM} have indicated that it is important to understand
neutrino cross section in the few GeV region. But we do not have
precise neutrino data to understand this region very well.
Thus, good modeling of neutrino cross sections at low energies 
becomes very crucial for the precise next generation neutrino 
oscillations experiments. In few GeV region, there are three types
of neutrino interactions, quasi-elastic, resonance, and inelastic
scattering. It is very challenging to disentangle each contribution 
separately, especially, resonance scattering vs deep inelastic
scattering (DIS) contribution, due to large non-perturbative QCD corrections
to the DIS contribution.

Our approach is that we describe these processes in terms of quark-parton
model using the precise charged-lepton scattering data. It is then
simpler to convert charged-lepton scattering cross section into neutrino
cross section. In a previous report~\cite{nuint02}, we showed that
our effective leading order model
using an improved scaling variable  $\xi_w$ describes
all deep inelastic scattering charged lepton-nucleon scattering data
including resonance data (SLAC/BCDMS/NMC/HERA/JLab)~\cite{slac,jlab} 
from very high $Q^2$ to very low $Q^2$ (down to photo-production region), 
as well as CCFR neutrino data~cite{yangthesis,rccfr}.

Our proposed scaling variable, $\xi_w$ is derived on following basis.
Using energy momentum conservation, the factional momentum, $\xi$ 
carried by a quark in a proton target of mass $M$ 
can be obtained as follows;
\begin{eqnarray}
 \xi &=& \frac{2xQ^{'2}}{Q^{2}(1+\sqrt{1+(2Mx)^{2}/Q^2})},
\end{eqnarray}
where
\begin{eqnarray}
2Q^{'2}  &=& [Q^2+M_f{^2}-M_i{^2}]  \nonumber \\
         &+& \sqrt{(Q^2 + M_f{^2}-M_i{^2})^2+4Q^2(M_i{^2}+P_{T}^{2})}\nonumber 
\end{eqnarray}
Here $M_i$ is the initial quark mass with average initial
transverse momentum $P_T$ and $M_f$ is the mass of the
quark in the final state.  This expression for $\xi$ 
was previously derived~\cite{gp} for the case of $P_T=0$. 
Assuming $M_i=0$ we use following variable in our model;
\begin{eqnarray}
\label{eq:xi}
\xi_w &=& \frac{2x(Q^2+M_f{^2}+B)}
        {Q^{2} [1+\sqrt{1+(2Mx)^2/Q^2}]+2Ax},
\end{eqnarray}
here, $M_f =0$ except charm-production case ($M_f$=1.5 GeV) in neutrino scattering.
The parameter $A$ 
accounts for the higher order (dynamic higher twist) QCD terms 
in the form of an enhanced target mass term (the effects of the proton
target mass are already taken into account using the exact form
in the denominator of $\xi_w$ ). The parameter
$B$ accounts for the initial state quark transverse
momentum and final state quark  effective $\Delta M_f{^2}$
 (originating from multi-gluon emission by quarks). 
This parameter allows that we could describe
the photoproduction limit (all the way down to $Q^{2}$=0).

A brief summary of our effective leading order (LO) model is given as follows;
\begin{enumerate}
 \item The GRV98 LO PDFs~\cite{grv98} are used to describe the $F_2$ data at high $Q^2$
	 region: The minimum $Q^2$ value of this PDFs is 0.8 GeV$^2$.
 \item  The scaling variable $x$ is replaced with the improved
        scaling variable $\xi_w$ in Eq.~\ref{eq:xi}. 
\item  Like earlier non-QCD based fits~\cite{DL} to low energy data, 
	we multiply all PDFs by $K$ factors, depending on the types
	of PDFs;
 	\begin{eqnarray}	
	\label{eq:kfac}
	 K_{sea}(Q^2) &=& \frac{Q^2}{Q^2 +C_s} \nonumber  \\
	 K_{valence}(Q^2) &=&[1-G_D^2(Q^2)] \nonumber  \\
	      &	\times & \left(\frac{Q^2+C_{v2}} 
		      {Q^{2} +C_{v1}}\right) 
	\end{eqnarray}	
	 where $G_D$ = $1/(1+Q^2/0.71)^2$ is the  proton elastic form factor.
	At low $Q^2$, $[1-G_D^2(Q^{2})]$
	is approximately $Q^2/(Q^2 +0.178)$, which is very close
	to our earlier fit result~\cite{nuint01}.
        These modifications were done in order to describe low $Q^2$
       data in the photoproduction limit, where 
	$F_{2}$ is related to the photoproduction cross section 
	according to
	\begin{eqnarray}
	     \sigma(\gamma p) = {4\pi^{2}\alpha_{\rm EM}\over {Q^{2}}}
	          F_{2}
	           = \frac{0.112 mb}{Q^2} F_2 
	\label{eq:photo} 
	\end{eqnarray}
 \item We freeze the evolution of the GRV98 PDFs at a
	value of $Q^2=0.80$. Below this $Q^2$, $F_2$ is given by;
	\begin{eqnarray}
	     F_2(x,Q^2<0.8) = K(Q^2) \times F_2(\xi,Q^2=8) 
	\end{eqnarray}

 \item Finally, we fit to the effective GRV98 LO PDFs $(\xi_w)$ 
       using the DIS data (SLAC/BCDMS/NMC/H1).
        We obtained excellent fitting results;
        $A$=0.419, $B$=0.223, and  $C_{v1}$=0.544, $C_{v2}$=0.431,
        and $C_{sea}$=0.380 and $\chi^{2}/DOF=$ 1235/1200.
	Because of the $K$ factors to the PDFs, we find that
        the GRV98 PDFs need to be increased by $N$=1.011.

\end{enumerate}

In the previous analysis, the structure functions data
	were corrected for the BCDMS systematic error shift and for
	the relative normalizations between the SLAC, BCDMS, NMC
	and H1 data. The deuterium data were corrected
	for nuclear binding effects~\cite{highx,nnlo}. 
	We also included charm production contribution 
	using the photon-gluon fusion
	model in order to fit the very high $\nu$ HERA data. This was not
	necessary for any of the low energy comparisons but was only required
	to describe the highest $\nu$ HERA $F_2$ and photoproduction data. 
	since the GRV98 PDFs did not include 
	the charm sea for $Q^{2}>0.8$ GeV$^2$.

Performance of this effective LO model was very good 
in describing various DIS data
from high $Q^2$ to low $Q^2$ region. Predictions for photo-production data
on proton and deuteron ($Q^2=0$ limit) showed good agreement.
Furthermore, this model showed a reasonable description 
of the average value of $F_2$
for SLAC and Jefferson resonance data~\cite{jlab} ($Q^2$ down to 0.07).

In this report, we improve our effective LO model by treating
low $Q^2$ corrections 
($K$ factor in Eq.~\ref{eq:kfac}) for $u$, $d$ valence and sea quarks separately.
Since the predictions of our previous model showed good agreements
with photoproduction data, we now include these data to get a better
constraint on our model, instead of predicting.
The fitted results using the improved model
are given as follow: $A$=0.538, $B$=0.305, $C_{v1d}$=0.202,
$C_{v1u}$=0.291, $C_{v2d}$=0.255, $C_{v2u}$=0.189, 
$C_{s1d}$=0.621, $C_{s1u}$=0.363, and $fPDF$=1.015,
and $\chi^{2}/DOF=$1874/1574.

\begin{figure}
\centerline{\psfig{figure=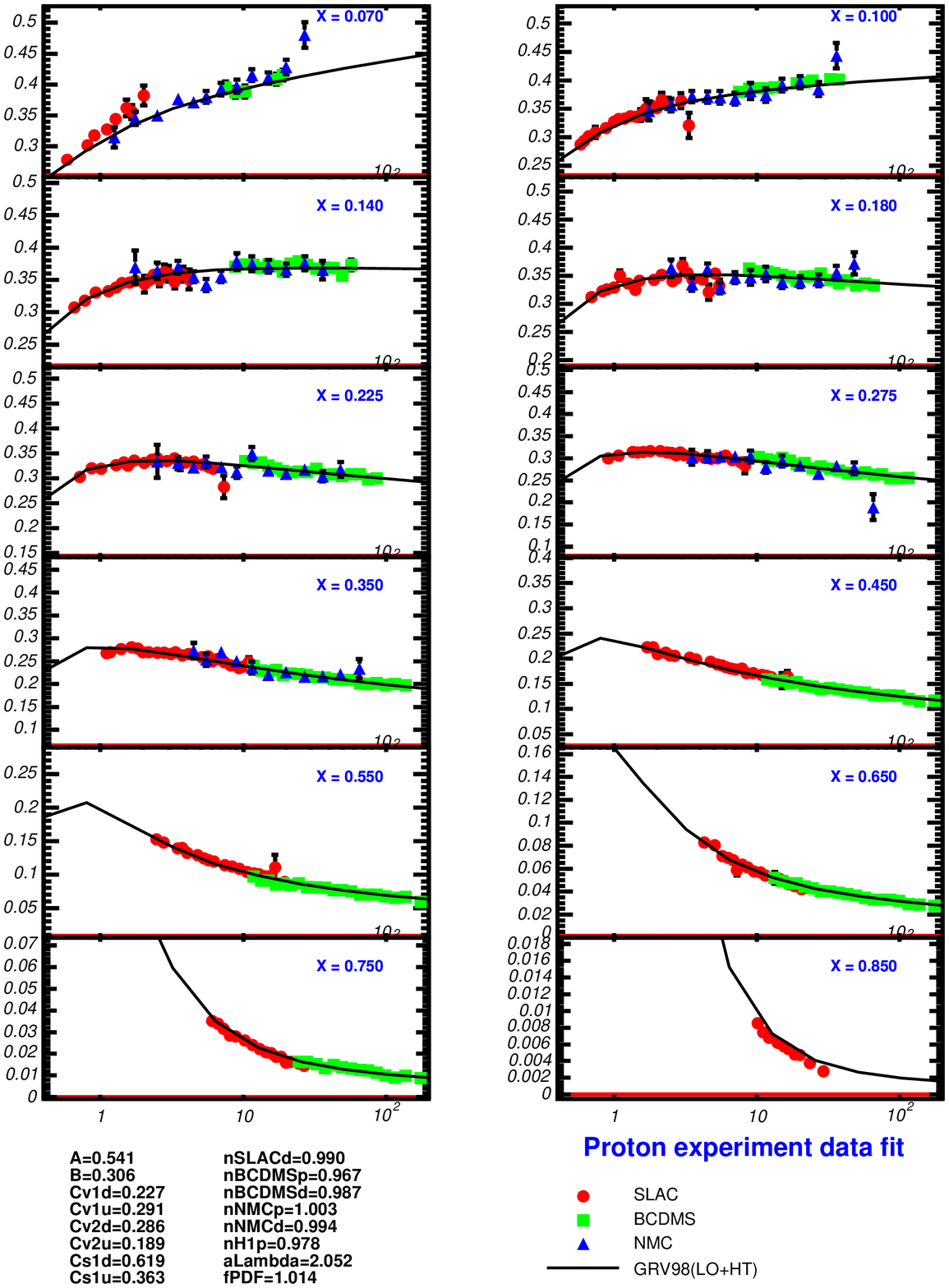,width=3.0in,height=3.5in}}
\centerline{\psfig{figure=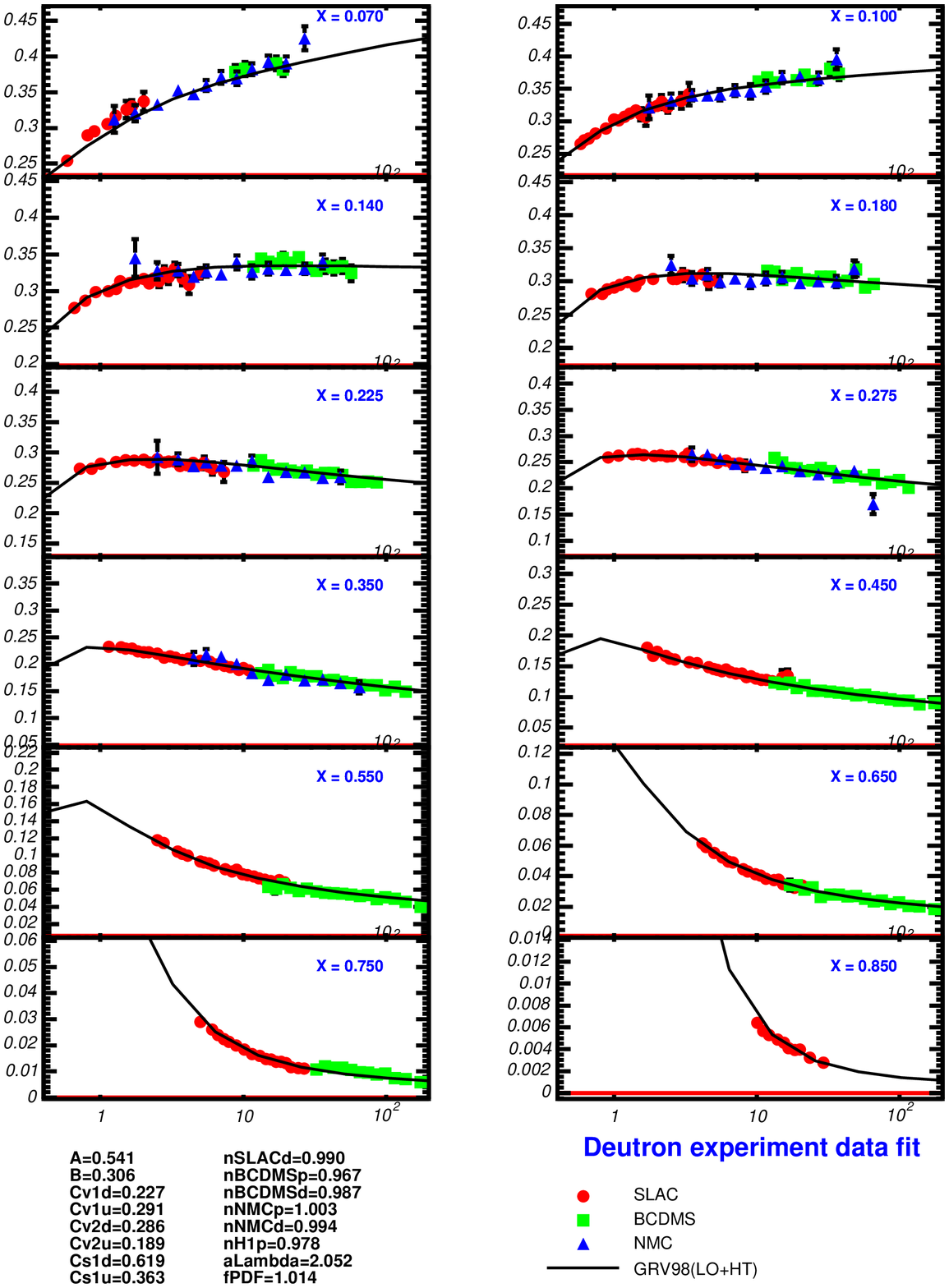,width=3.0in,height=3.5in}}
\caption{Our fit results to the effective LO model 
is compared to the $F_2$ data (SLAC, BCDMS, NMC) 
at high $x$:[top] $F_2$ proton, [bot] $F_2$ deuteron.}
\label{fig:f2fit_highx}
\end{figure}

\begin{figure}
\centerline{\psfig{figure=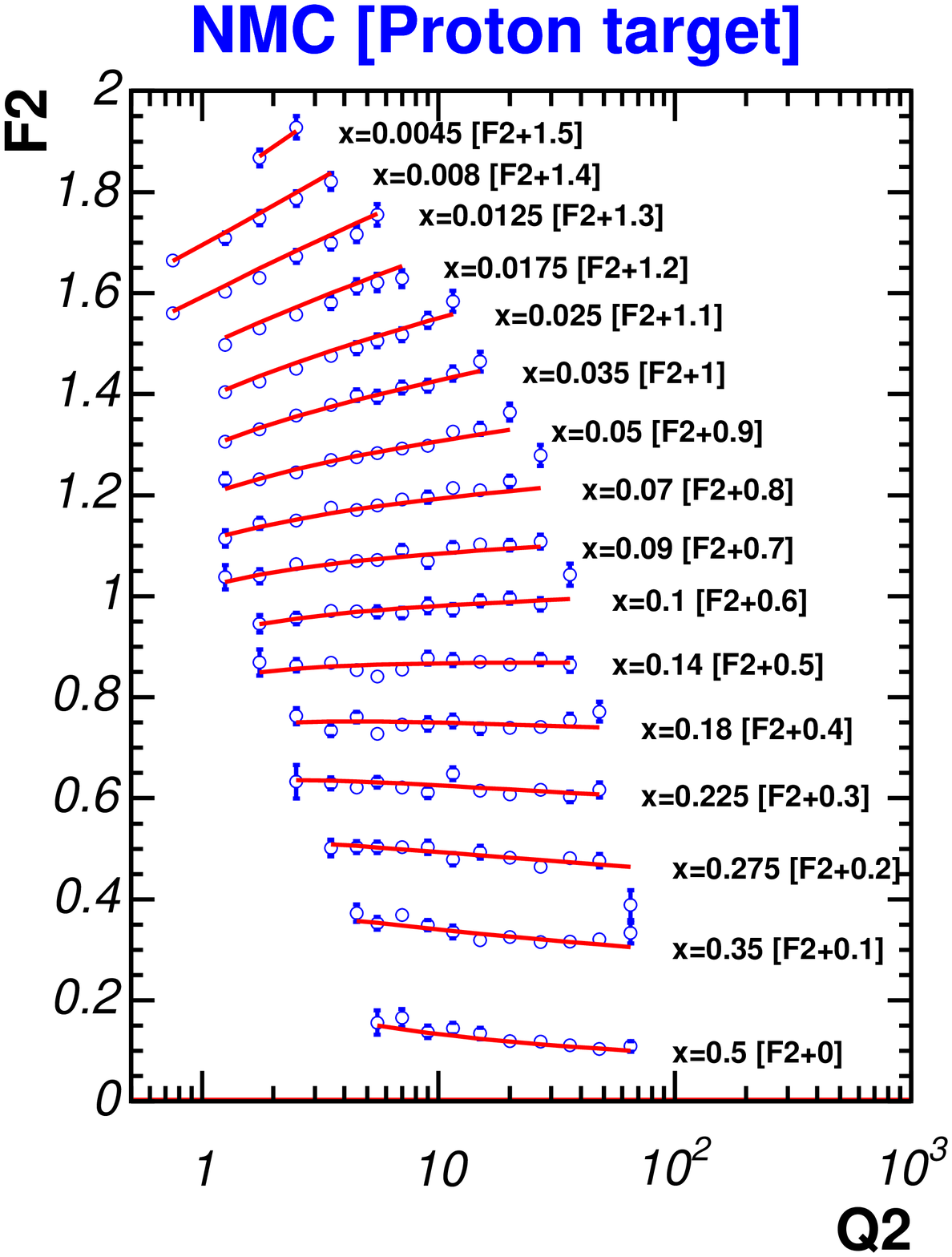,width=3.0in,height=3.5in}}
\centerline{\psfig{figure=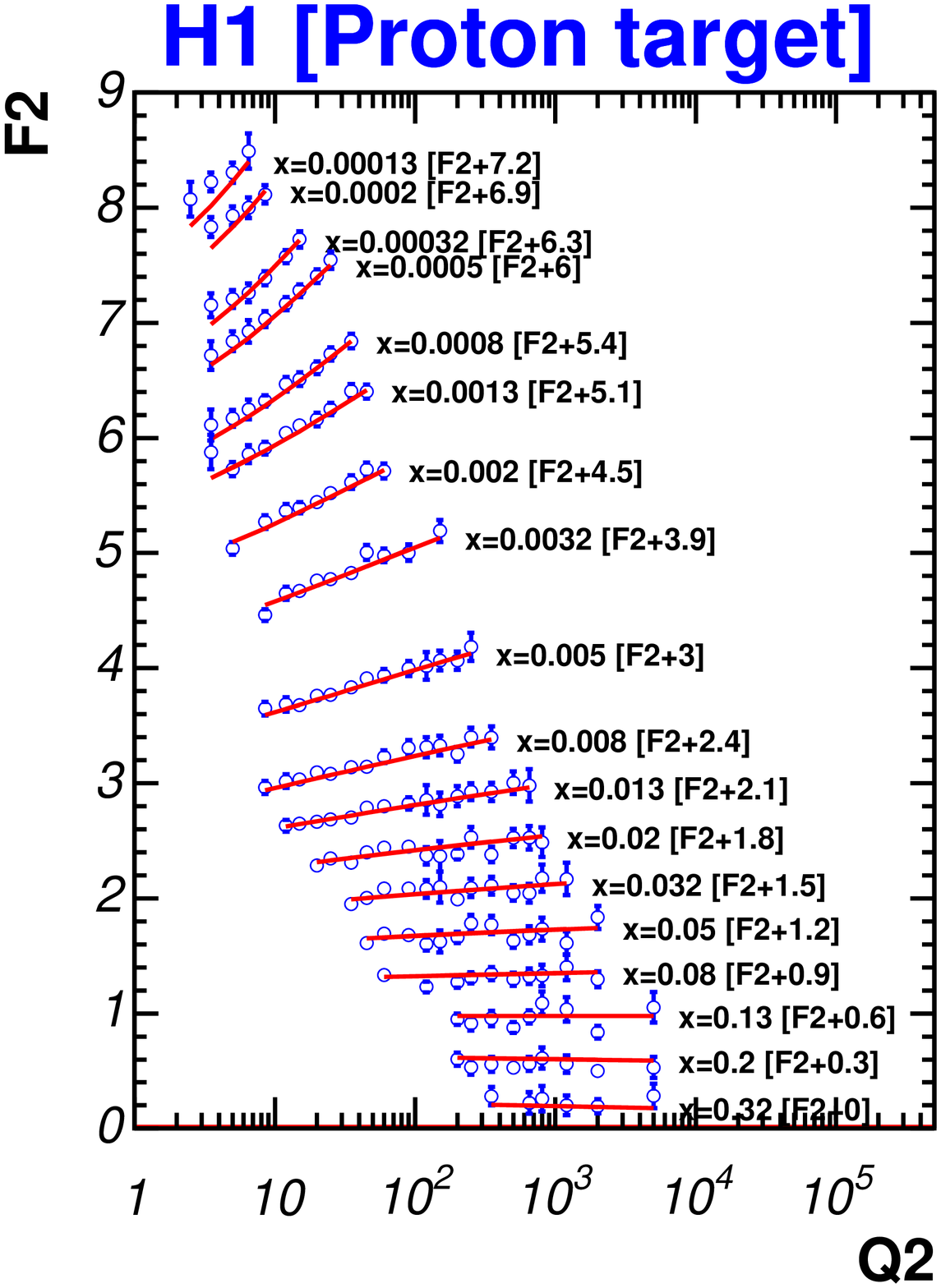,width=3.0in,height=3.5in}}
\caption{Our fit results to the effective LO model 
are compared to the $F_2$ data at low $x$: [top] NMC $F_2$ data, 
[bot] H1 $F_2$ data.}
\label{fig:f2fit_lowx}
\end{figure}

\begin{figure}
\centerline{\psfig{figure=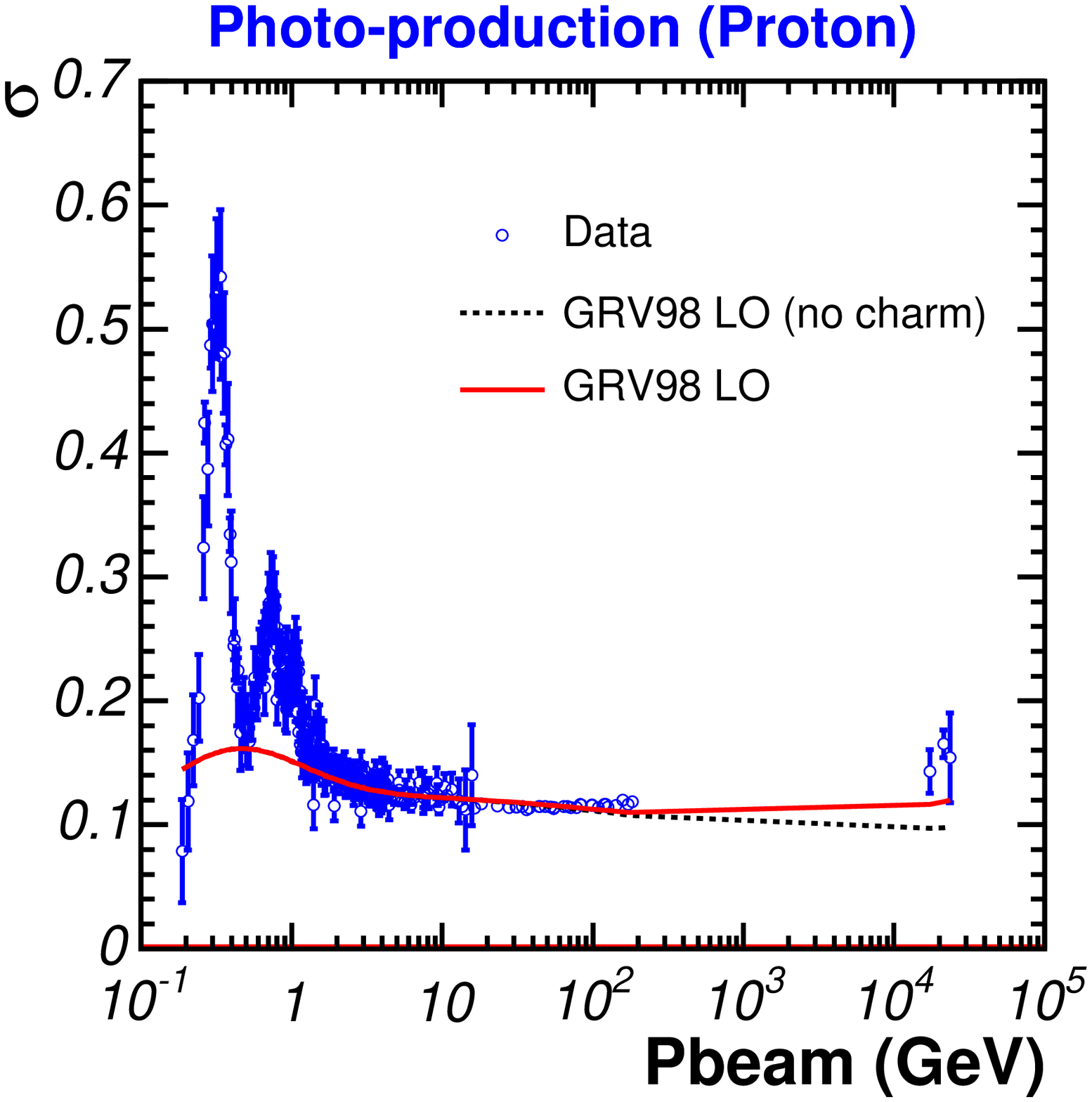,width=3.0in}}
\centerline{\psfig{figure=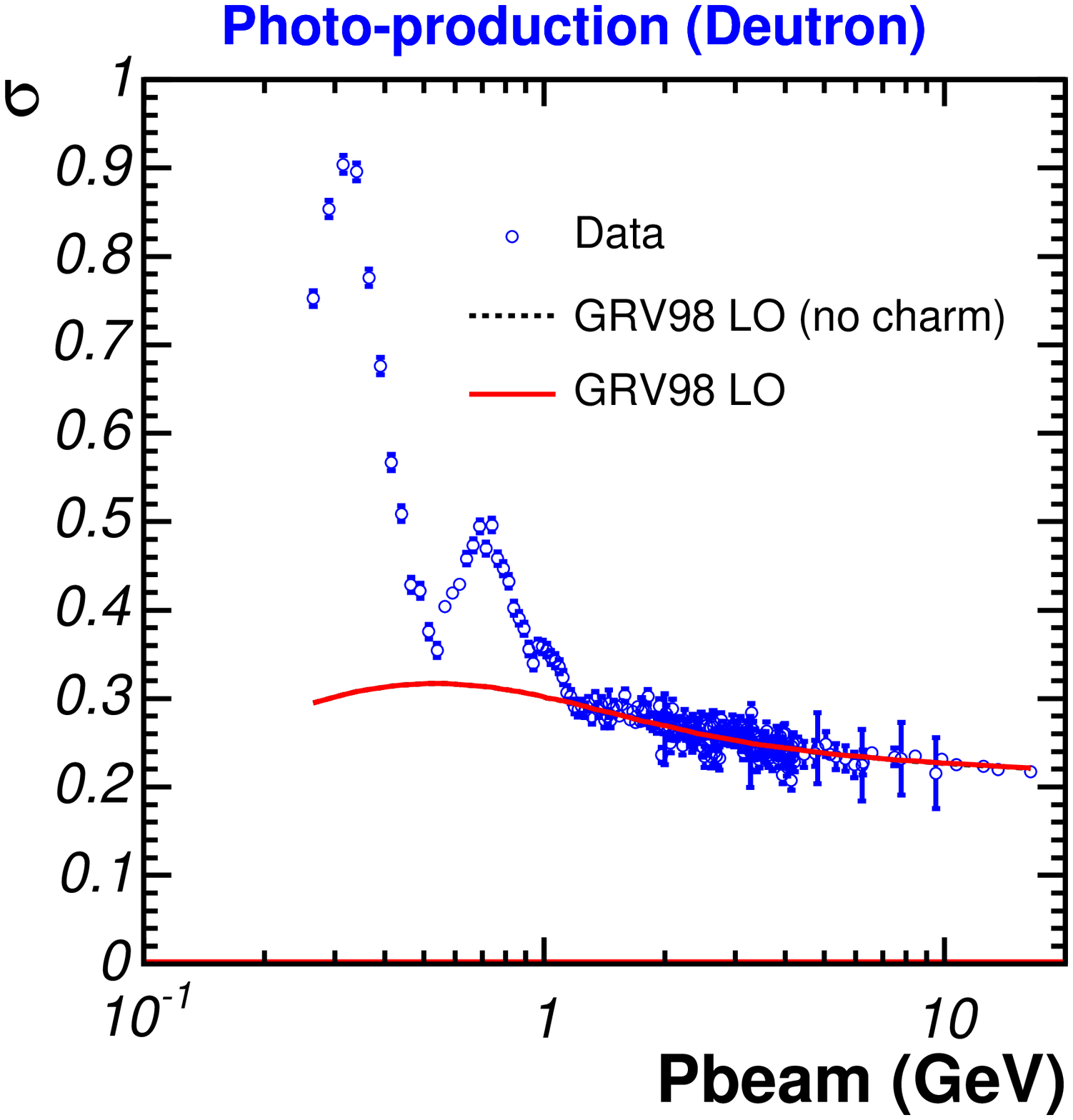,width=3.0in}}
\caption{Our fit results  to the effective LO model 
are compared to the photoproduction data ($Q^2=0$ limit)
[top] proton, [bot] deuteron.
The dotted line do not include charm contribution from gluon fusion process}
\label{fig:photo}
\end{figure}

Comparison of all DIS $F_2$ data and our fits
are shown in Figure~\ref{fig:f2fit_highx} and ~\ref{fig:f2fit_lowx}.
Our effective LO model describes the DIS $F_2$ data at low $x$ 
and high $x$ regions as well. This model also fits
the photoproduction data on proton and deuteron targets very well,
as shown in Figure~\ref{fig:photo}.

\begin{figure}[t]
\centerline{\psfig{figure=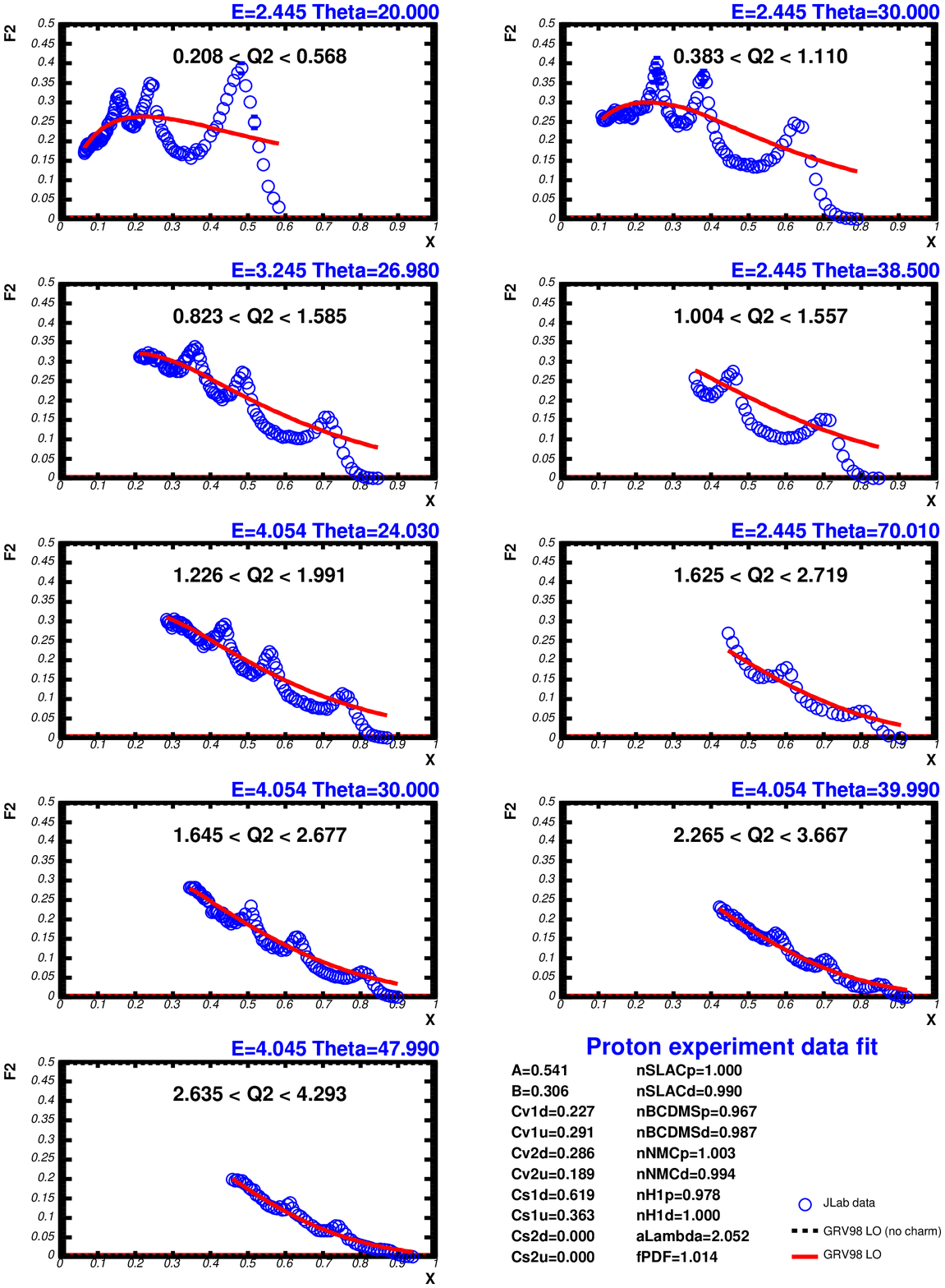,width=3.0in,height=3.5in}}
\centerline{\psfig{figure=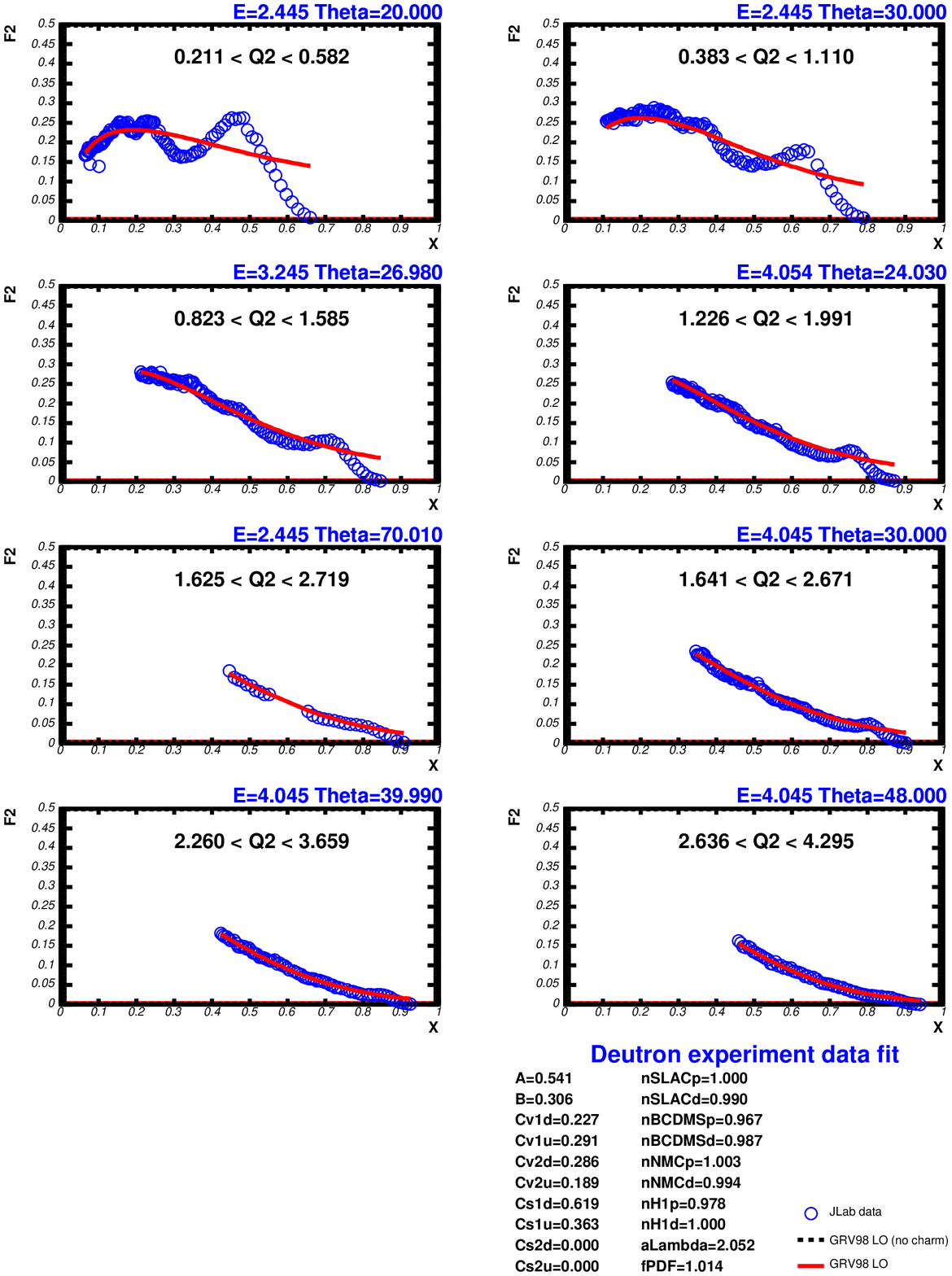,width=3.0in,height=3.5in}}
\caption{ Comparisons of the resonance data 
and the predictions of our effective LO model:
[top] proton, [bot] deuteron.}
\label{fig:res}
\end{figure}

Comparisons of predictions using our model to the resonance data~\cite{jlab}
which are not included in the fit is shown in Figure~\ref{fig:res}.
Based on duality argument~\cite{bloom},
our model provides a reasonable description of the average value of $F_2$ data
in the resonance region (down to $Q^{2}=0.07$). 

\begin{figure}
\centerline{\psfig{figure=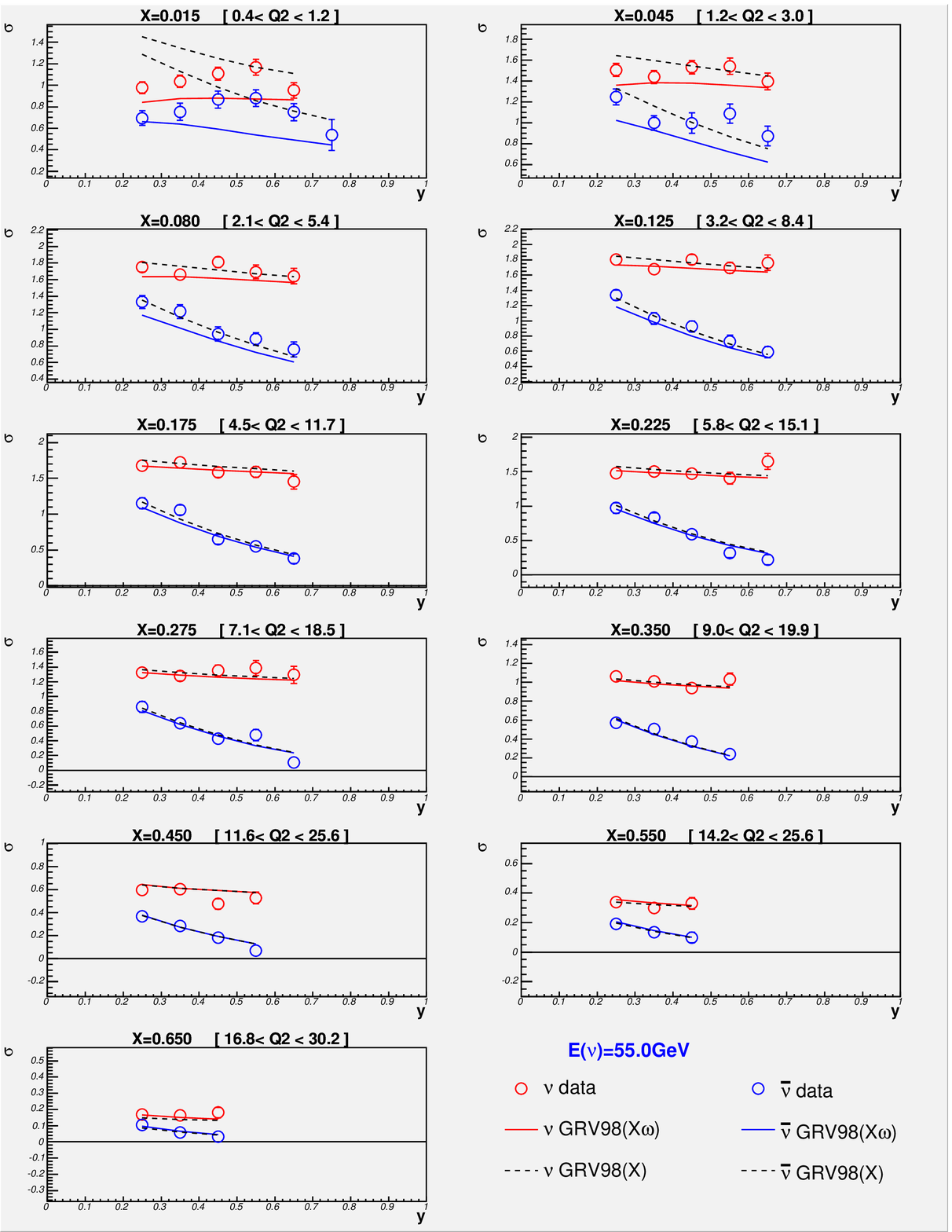,width=3.0in,height=3.5in}}
\centerline{\psfig{figure=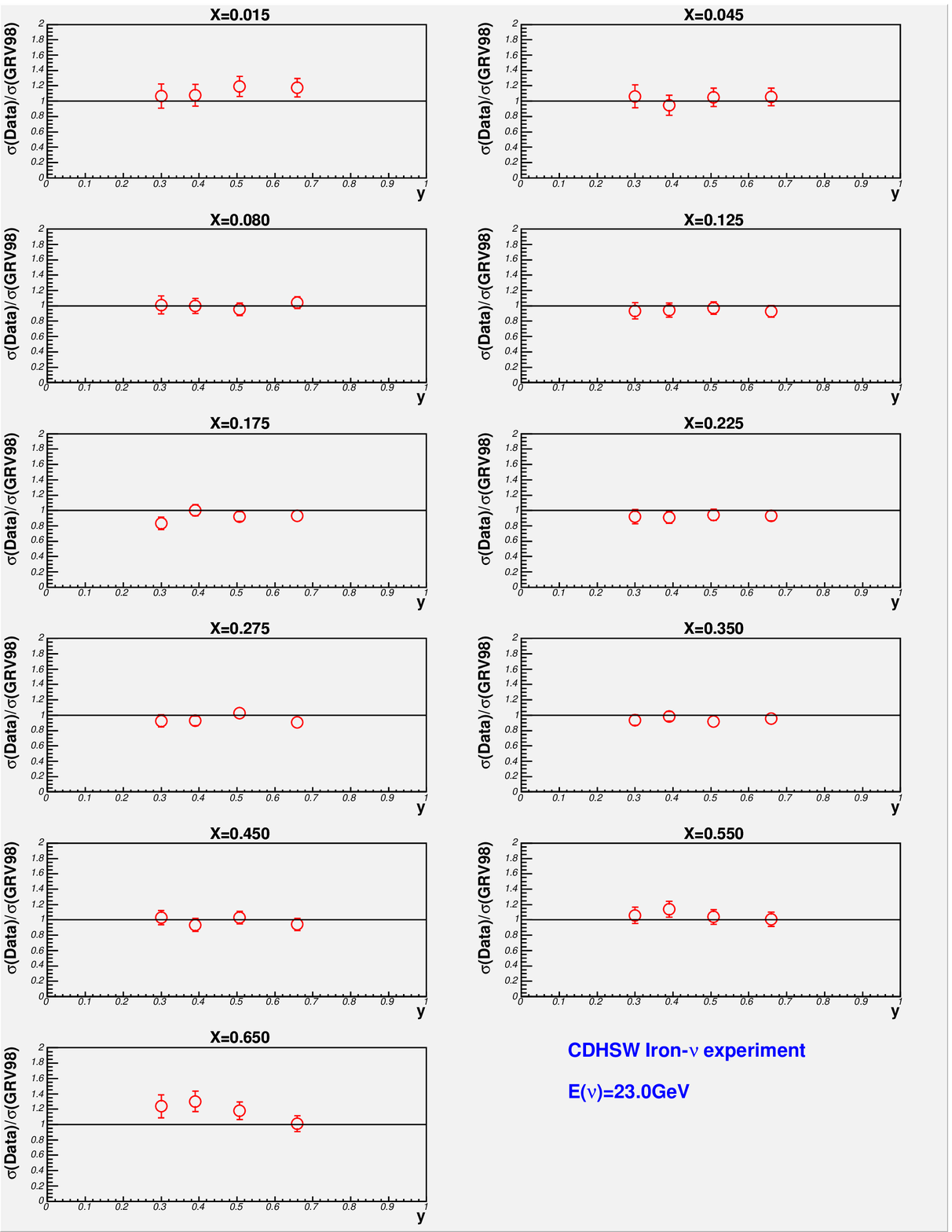,width=3.0in,height=3.5in}}
\caption{[top] Comparisons of the CCFR neutrino data and the predictions
(solid line) of our effective LO model at 55 GeV, the dotted line
using the default GRV98 PDFs: 
[bot] Ratio of the CDHSW neutrino data and the predictions
of our effective LO model at 23 GeV.}
\label{fig:neutrinoD}
\end{figure}

In addition to vector structure function, we have an axial vector structure
function in neutrino scattering.
At the $Q^2=0$ limit, the vector structure function
should go to zero, 
while the axial-vector part has a finite contribution. 
At high $Q^2$, these two structure
functions are expected to be same. Thus, it is important
to understand the axial vector contribution at low $Q^2$
using the low energy neutrino data.
As a preliminary step, we compare the CCFR~\cite{yangthesis,rccfr}
and CDHSW energy~\cite{cdhsw}
neutrino data with our model, assuming that the vector contribution 
is same as the axial vector contribution.
Figure~\ref{fig:neutrinoD} show comparisons of our model
to the high energy CCFR and CDHSW neutrino charged-current
cross section data.
In this comparison we correct for nuclear effects in iron.
The structure function $2xF_{1}$
is  obtained by using the $R_{world}$ fit from reference~\cite{slac}.
Our predictions show very good agreement
with these neutrino data on iron.

In order to have a full description of all charged current
processes, the contribution from quasi-elastic
scattering must be added separately at $x=1$.   
The best prescription is to
use our model in the region above the first resonance (above $W$=1.35 GeV) 
and add the contributions  from
quasi-elastic and first resonance ($W$=1.23 GeV) separately.
This is because the  $W$=$M$ and  $W$=1.23 GeV regions are dominated by
one and two isospin states, and the amplitudes for neutrino
versus electron scattering are related via Clebsch-Gordon rules~\cite{rs}
instead of quark charges (also the $V$ and $A$
couplings are not equal at low $W$ and $Q^2$).
In the region of higher mass resonances
      (e.g. $W$=1.7 GeV) there is a significant
contribution from the deep-inelastic continuum which is not
well modeled by the existing fits~\cite{rs} to
neutrino resonance data (and using our modified
PDFs should be better).
For nuclear targets, nuclear corrections~\cite{nuint01} must also be applied.
Recent results from Jlab indicate that the Fe/D ratio in the
resonance region is the same as the Fe/D ratio 
from DIS data for the same value of $\xi$ (or $\xi_w$).

We are currently working on constraining the low $Q^2$ axial vector
contribution using low energy CDHSW and CHORUS~\cite{chorus} data.
The form of the fits we plan to use is motivated
by the Adler sum rule~\cite{adler} for the axial vector contribution as follows:
\begin{eqnarray}	
 K_{sea-ax}(Q^2) &=& \frac{Q^2+C_{2s-ax}}{Q^2 +C_{1s-ax}} \nonumber  \\
 K_{valence}(Q^2) &=&[1-F_A^2(Q^2)] \nonumber  \\
      &	\times & \left(\frac{Q^2+C_{2v-ax}} 
	      {Q^{2} +C_{1v-ax}}\right), 
\end{eqnarray}	
where $F_{A}(Q^2) = -1,267/(1+Q^2/1.00)^2$.

In addition, we plan to implement a modulating 
function~\cite{bodek,close} A(W,$Q^{2}$) to improve modeling
in the resonance region by including (instead of predicting)
the resonance data. Although a description of the average cross section 
in the resonance region is sufficient for most neutrino experiments,
because of the effects of experimental resolution 
and Fermi motion for nuclear targets.


\begin{thebibliography}{9}

\bibitem{ATM}  S. Fukuda {\em et al.}, Phys.
Rev. Lett. {\bf 85}, 3999 (2000); T. Toshito,   hep-ex/0105023.

\bibitem{nuint02} A. Bodek and U. K. Yang, hep-ex/0308007
\bibitem{nuint01} A. Bodek and U. K. Yang, hep-ex/0203009
Nucl.Phys.Proc.Suppl.112:70-76,2002

\bibitem{slac} L. W. Whitlow {\it et al.}  (SLAC-MIT), Phys. Lett. B
{\bf 282}, 433 (1995);
A. C. Benvenuti {\it et al.} (BCDMS), Phys. Lett. B{\bf237}, 592 (1990);
 M. Arneodo {\it et al.}  (NMC), Nucl. Phys. B{\bf 483}, 3 (1997).
\bibitem{jlab}
C. Keppel, Proc. of the Workshop on Exclusive Processes 
at High $P_T$, Newport News,
VA, May (2002).]

\bibitem{yangthesis}
U. K. Yang, Ph.D. thesis, Univ. of Rochester, UR-1583 (2001).
 http:/hep.uchicago.edu/\\ $\sim$ukyang/neutrino/thesis.ps.
\bibitem{rccfr} U. K. Yang {\it et al.}(CCFR), Phys. Rev. Lett. 
{\bf 87}, 251802 (2001).


\bibitem{gp} H. Georgi and H. D. Politzer, Phys. Rev. D{\bf 14}, 1829 
(1976); R. Barbieri {\it et al.},  Phys. Lett. B{\bf
64}, 171 (1976), and  Nucl. Phys. B{\bf
117}, 50 (1976); J. Pestieau and J. Urias, Phys.Rev.D{\bf 8}, 1552 (1973)

\bibitem{grv98} M. Gluck, E. Reya, A. Vogt, Eur. Phys. J {\bf C5}, 461 (1998).

\bibitem{DL} A. Donnachie and P. V. Landshoff, Z. Phys. C {\bf61}, 139 (1994);
B. T. Fleming {\it et al.}(CCFR), Phys. Rev. Lett. {\bf 86}, 5430 (2001).


\bibitem{highx} U. K. Yang and A. Bodek, Phys. Rev. Lett. {\bf
82}, 2467 (1999).
%
\bibitem{nnlo} U. K. Yang and A. Bodek, Eur. Phys. J. C{\bf
13}, 241 (2000).
%

\bibitem{cdhsw}
P. Berge {\em et al.} (CDHSW), Zeit. Phys. {\bf C49}, 607 (1991).

\bibitem{rs} D. Rein and L. M. Sehgal, Annals Phys. {\bf 133}
79 (1981); and  R. Belusevic and D. Rein,
Phys. Rev. D {\bf 46}, 3747 (1992).

%
\bibitem{bloom}E. D. Bloom and F. J. Gilman, Phys. Rev. Lett. {\bf
25}, 1140 (1970).
%

\bibitem{chorus} R. Oldeman, Proc. of 
30th International Conference on High-Energy Physics (ICHEP 2000), 
Osaka, Japan, 2000.

\bibitem{adler} S. Adler, Phys. Rev. {\bf 143}, 1144 (1966);
F. Gillman, Phys. Rev. {\bf 167}, 1365 (1968).
%

\bibitem{bodek} A. Bodek {\it et al.}, Phys. Rev. D{\bf 20}, 1471 (1979).

\bibitem{close} S. Stein {\it et al.}, Phys. Rev. D{\bf 12}, 1884 (1975);
 K. Gottfried,  Phys. Rev. Lett. {\bf
18}, 1174 (1967).

\end{thebibliography}
\end{document}